\documentclass[11pt,a4paper]{emulateapj}
\usepackage{amsmath}
\usepackage{multirow}

\begin{document}

\title{Chandra Observation of Abell 1142: A Cool-Core Cluster Lacking a Central Brightest Cluster Galaxy?}

\author{Yuanyuan Su\altaffilmark{$\dagger$1,2}}
\author{David A. Buote\altaffilmark{2}}
\author{Fabio Gastaldello\altaffilmark{3}}
\author{Reinout van Weeren\altaffilmark{1}}
\affil{$^1$Harvard-Smithsonian Center for Astrophysics, 60 Garden Street, Cambridge, MA 02138, USA}
\affil{$^2$Department of Physics and Astronomy, University of California, Irvine, 4129 Frederick Reines Hall, Irvine, CA 92697, USA}
\affil{$^3$INAF-IASF-Milano, Via E. Bassini 15, I-20133 Milano, Italy}

\altaffiltext{$\dagger$}{Email: yuanyuan.su@cfa.harvard.edu}

\begin{abstract}

Abell~1142 is a low-mass galaxy cluster at low redshift containing two comparable Brightest Cluster Galaxies (BCG) resembling a scaled-down version of the Coma Cluster. 
Our {\sl Chandra} analysis reveals an X-ray emission peak, roughly 100\,kpc away from either BCG,  which we identify as the cluster center. 
The emission center manifests itself as a second beta-model surface brightness component distinct from that of the cluster on larger scales. The center is also substantially cooler and more metal rich than the surrounding intracluster medium (ICM), which makes Abell~1142 appear to be a cool core cluster.
The redshift distribution of its member galaxies indicates that Abell~1142 may contain two subclusters with each containing one BCG.   
The BCGs are merging at a relative velocity of $\approx$1200 km\,s$^{-1}$. This ongoing merger may have shock-heated the ICM from $\approx$\,2 keV to above 3 keV, which would explain the anomalous $L_{\rm X}$--$T_{\rm X}$ scaling relation for this system.
This merger may have displaced the metal-enriched ``cool core" of either of the subclusters from the BCG.
The southern BCG consists of three individual galaxies residing within a radius of 5 kpc in projection.  
These galaxies should rapidly sink into the subcluster center due to the dynamical friction of a cuspy cold dark matter halo.

\end{abstract}

\section{\bf Introduction}

X-ray observations reveal two populations of galaxy clusters: cool-core clusters (CC) and non-cool-core clusters (NCC). CC clusters possess a sharp X-ray emission peak with a positive radial temperature gradient owing to the radiative cooling of its dense core (e.g., Molendi \& Pizzolato 2001; Sanders et al.\ 2004; Sanderson et al.\ 2006). The metal distribution in the intracluster medium (ICM) of CC clusters peaks at cluster center, which further enhances the cooling process (e.g., De Grandi \& Molendi 2001, Ettori et al.\ 2015). In contrast, the gaseous, thermal, and chemical distributions of NCC clusters are relatively homogenous. Cosmological simulations also reproduced the metallicity and temperature profiles of CC clusters, which are steeper than those of NCC clusters (e.g., Rasia et al.\ 2015).

A fundamental question is whether CC and NCC are formed intrinsically differently or whether any cluster can transform into either type over cosmic time. 
Cavagnolo et al.\ (2009) found that the central entropy of galaxy clusters is bimodal with two distinct peaks at 15 keV cm$^2$ and 150 keV cm$^2$ respectively. 
Non-gravitational preheating may have increased the entropy of NCC clusters to above 300 keV cm$^2$, and they never had enough time to develop a cool core.
This difference may also be set by different levels of conduction. A critical entropy threshold (30--50 keV\,cm$^2$) may exist, only below which can conduction be halted and a cool core form (Voit et al.\ 2008; Guo et al.\ 2008).

Meanwhile, many scenarios have been proposed arguing against a primordial origin of the CC/NCC dichotomy. 
Heating from active galactic nuclei (AGN) and their associated bubbles may counteract radiative cooling at the cluster center (e.g., Fabian 2012; Guo \& Mathews 2010; McNamara \& Nulsen 2007). 
If CC and NCC clusters are distinguished by their central AGN activity, it would not explain why CC clusters are more relaxed than NCC clusters.
Yet, observations show that most CC clusters appear to be more symmetric than NCC clusters (Buote \& Tsai 1996; Sanderson et al.\ 2009) 
These results favor major mergers as the origin of NCC clusters. 
Major mergers may have wiped out nascent cool cores in NCC clusters, while CC clusters experience only minor mergers and preserve their cool cores (Henning et al.\ 2009; Burns et al.\ 2008).   
In principle, all NCC clusters will eventually evolve into CC clusters given enough time to relax (McGlynn \& Fabian 1984; Buote 2001a). 
Nevertheless, Motl et al.\ (2004) proposed that mergers may even transform NCC clusters into CC clusters with cold gas supplied by infalling sub-groups.

As the largest gravitationally bound objects in the Universe, galaxy clusters are able to retain the bulk of metals synthesized by supernova. 
The metal abundance ratios in the ICM and their spatial distributions provide important clues to the chemical enrichment history and the formation process of galaxy clusters.      
Previous X-ray observations of groups and clusters indicate a centrally-peaked Fe profile coupled with a relatively homogenous distribution of O and Si (e.g., Finoguenov et al.\ 2000; Bohringer \& Werner 2010; Sanderson \& Ponman 2010; Buote 2001b). This implies that $\alpha$-elements have been supplied during the pre-enrichment phase by young massive stars through Type II supernova (SNII), while most Fe has been ejected into the ICM more recently by the central BCG through Type Ia supernova (SNIa). However, a few studies using {\sl Chandra} or {\sl XMM-Newton} observations report a Si distribution just as centrally-peaked as Fe (e.g., Buote et al.\ 2003; Million et al.\ 2011; Simionescu et al.\ 2010).  
The cluster center may be more enriched by SNII than previously thought, which challenges our understanding of the history of galaxy clusters.  
Elkholy et al.\ (2015) measured a larger scatter in the central Fe mass among low-mass clusters relative to high-mass clusters. They interpret this as more frequent mergers (mixing) of sub-groups in larger clusters, resulting in a more constant metal content.
Metal-rich gas from an infalling galaxy or sub-group has been seen to reach the cluster center in some simulations (e.g., Cora 2006), which explains the observed SNII products at the cluster center.
It is unclear whether the central metal excess in the ICM is due to the BCG or due to the infall of metal-rich gas.
An ideal test is to study the central metal content of a galaxy cluster without a BCG at its X-ray center. 
Unfortunately, most such systems are NCC clusters that lack a metal abundance gradient, likely having had their central metal peak destroyed by cluster mergers (e.g., Matsushita 2011; De Grandi \& Molendi 2001).

\begin{figure}[h]
   \centering
    \includegraphics[width=0.475\textwidth]{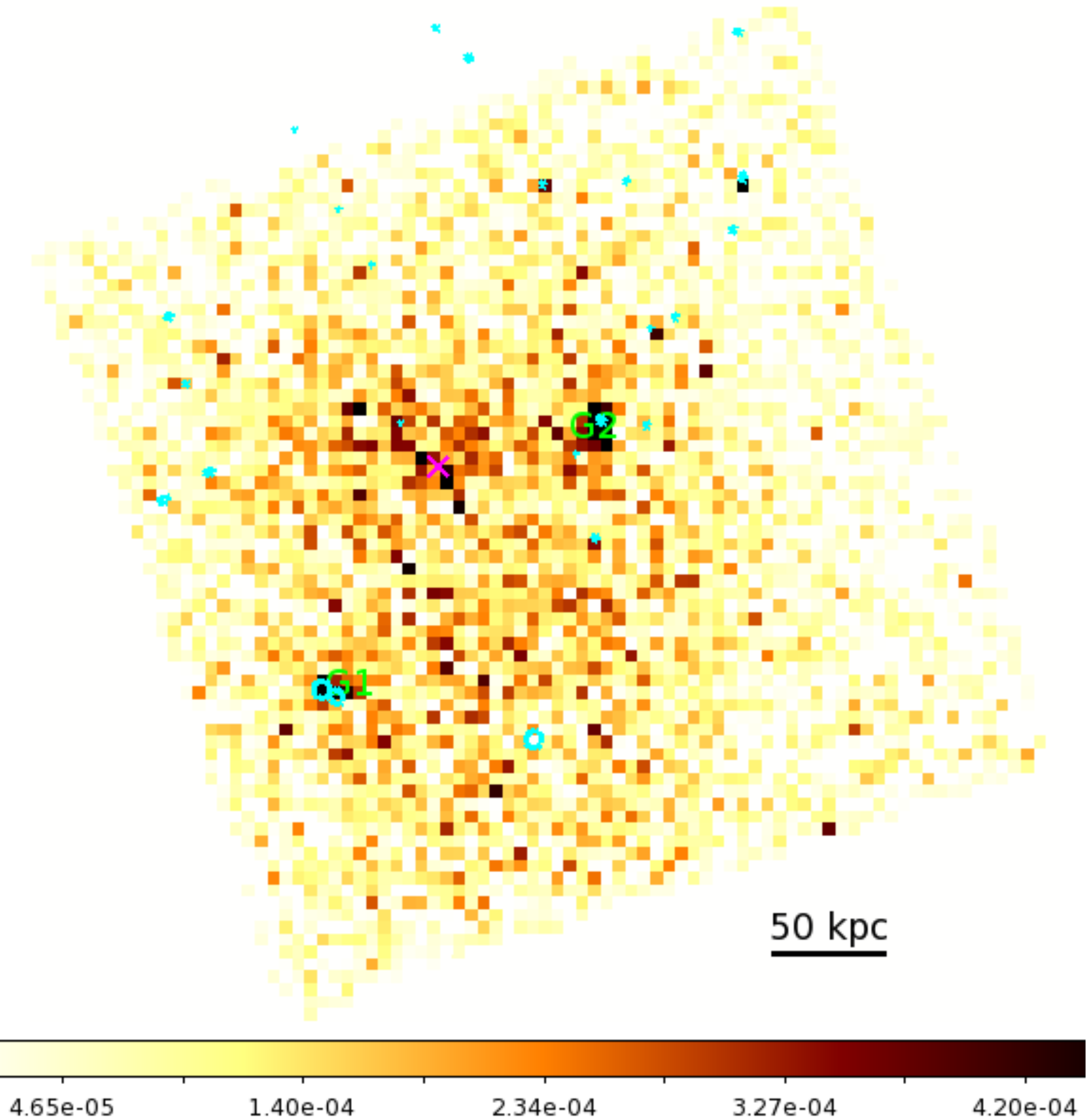}
     \includegraphics[width=0.475\textwidth]{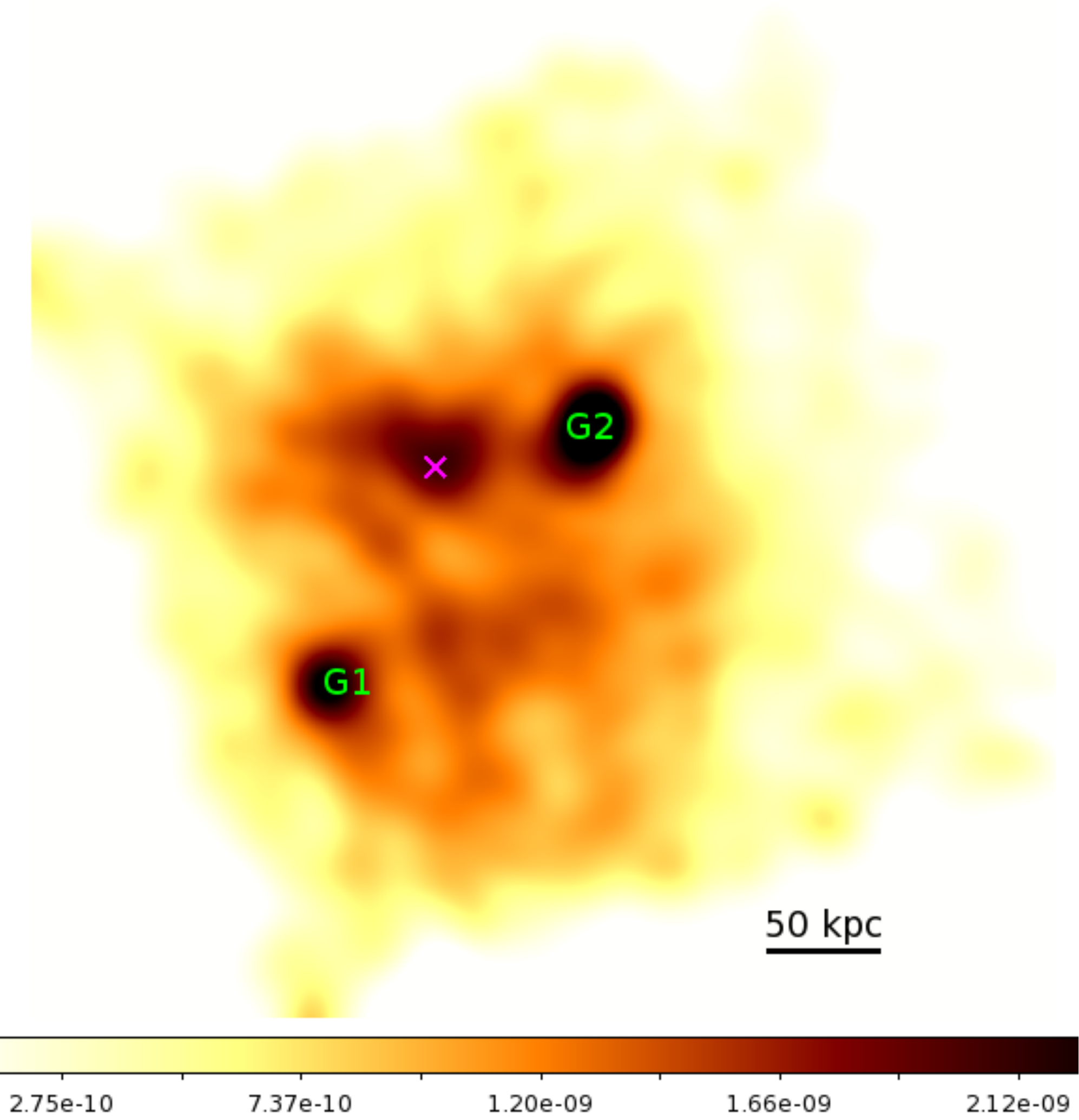}
    
\figcaption{\label{fig:A1142}27 ksec exposure-corrected and blank-sky background subtracted {\sl Chandra} ACIS-S image of Abell~1142 and in the energy band of 0.5--4.0 keV. {\it top}: The image was binned by 16$\times$16 pixels in the unit of counts/s. Positions of point sources were marked in cyan ellipticals. {\it bottom}: adaptively smoothed in the unit of photon\,cm$^{-2}$\,s$^{-1}$.
Locations of G1 and G2 are indicated. The peak of cluster gas emission is indicated by magenta cross.}
\end{figure}

Abell~1142, presented in this paper, is a poor cluster cataloged in the Northern {\sl ROSAT} All-Sky (NORAS) galaxy cluster survey (Bohringer et al.\ 2000) with an X-ray luminosity of $L_{\rm X}=2.8\times10^{43}$erg\,s$^{-1}$ in the 0.1--2.4 keV band.
Rather than having a definitive BCG,
it contains two bright galaxies (indicated by G1 and G2 in Figure~\ref{fig:A1142}). Lin \& Mohr (2004) take G1 (NGC~3492; $z$=0.036; galaxy pair) as the BCG, while Abdullah et al.\ (2011) take G2 (IC~664; $z$=0.034; S0) as the BCG. As demonstrated below, the ICM of Abell~1142 harbors a cool core detached from either BCG.
Abell~1142 provides us an opportunity to study a ``cool core" in the absence of a BCG as well as the role of BCG in the cluster enrichment process. In this work, we report our {\sl Chandra} analysis of Abell~1142.

We assume a luminosity distance of 152 Mpc for Abell~1142 based on a redshift of $z=0.035$ taken from the NASA/IPAC Extragalactic Database (NED). A cosmology with $H_0=70$ km s$^{-1}$ Mpc$^{-1}$,
$\Omega_{\Lambda}=0.7$ and $\Omega_m=0.3$ was adopted, corresponding to a linear scale of $1^{\prime} = 41.3$ kpc at the assumed distance. 
All uncertainty ranges are 68\% confidence intervals (1 $\sigma$), unless stated otherwise.

\section{\bf observations and data reduction}

\subsection{Chandra}
 
Abell~1142 was observed with {\sl Chandra} for 27 ksec on 2006-01-17 (ObsID: 06938). 
The aimpoint was on the back-side illuminated ACIS-S3 CCD.  The standard {\sl Chandra} data processing packages {\sl CIAO}~4.7 and {\sl CALDB}~4.6.9 were used to process and analyze the data. 
The observation was reprocessed from level 1 events using {\sl CIAO} tool {\tt repro}. CTI and time-dependent gain corrections were applied. 
Bad pixels, bad columns, and node boundaries were discarded.
Background flares beyond 3 $\sigma$ was filtered with the light curve filtering script {\tt lc\_clean}.  The entire 27 ksec observation was fairly quiescent. 
We identified point sources in the 0.3--7.0 keV energy band using {\tt wavdetect}; a 2.0 keV exposure map was supplied. The detection threshold was set at 10$^{-6}$ and the scales ranged from 1 to 8, in steps increasing by a factor of $\sqrt{2}$.

\subsubsection{\bf Imaging analysis}

We generated images in 8 energy bands: 0.5--0.8 keV, 0.8--1.2 keV, 1.2--1.5 keV, 1.5--2.0 keV, 2.0--2.5 keV, 2.5--3.0 keV, 3.0--3.5 keV, and 3.5--4.0 keV. We normalized these images with monochromatic exposure maps defined at the central energy of each band. For each image we subtracted an estimate of the background using the blank-sky fields available in the Chandra calibration database (CALDB). The background level was normalized by its count rate in 9.5 -- 12.0 keV relative to the observation. We filled point sources with pixel values interpolated from surrounding background regions using {\tt dmfilth}. A final 0.5--4.0 keV image was produced by adding all these 8 narrow band images. We further adaptively smoothed the image using {\tt dmimgadapt} as shown in Figure~\ref{fig:A1142}.

The distribution of the hot gas is very diffuse without a dominant central emission peak, but rather possesses multiple peaks of modest intensity.
The X-ray emission contours reveal three peaks (labeled as G1, G2, and X), offset by $\sim$100--200\,kpc with each other. Peaks G1 and G2 overlap with the BCGs G1 and G2 respectively, while Peak X corresponds to no big galaxy.  
The count rate (within 20 kpc) of Peak X (0.01 cts/s) is comparable those of Peak G1 (0.007 cts/s) and Peak G2 (0.012 cts/s). We adopt Peak X as the ``center" of the cluster for the following analyses since it lies roughly midway between G1 and G2 closer to the ICM center.

\subsubsection{\bf Spectrum analysis}

Individual spectra were extracted from four concentric circular regions centered on Peak X: 0--65$^{\prime\prime}$, 65-123$^{\prime\prime}$, 123--177$^{\prime\prime}$, and 177--295$^{\prime\prime}$.
Approximately the same number of net counts were contained in each region and they were just sufficient for spectral analysis. 
Instrument response files were generated for each spectrum using the CIAO tools {\tt mkwarf} and {\tt mkacisrmf}. 
The spectral fit was performed in the energy band of 0.5--7.0 keV with {\sl XSPEC} 12.7. C-statistic was used to measure goodness of the fit. The spectra were grouped to have at least 20 photons per energy bin in order to increase computational speed (Humphrey, Liu, \& Buote 2009; Su et al.\ 2015a). All four annuli were fitted simultaneously. 
Both the cluster emission and background components were modeled.

The X-ray background was approximated by {\tt apec$_{\rm LB}$}+{\tt phabs}$\times$({\tt apec$_{\rm MW}$}+pow$_{\rm CXB}$). An {\tt apec} thermal emission model represents the Local Bubble component; an additional {\tt apec} model is for the Milky Way emission (Smith et al.\ 2001), 
and a power law model {\tt pow} ($\Gamma=1.41$) for the unresolved cosmic X-ray background (De Luca \& Molendi 2004).
The temperatures of the Local Bubble and the Milky Way components were fixed at 0.08 keV and 0.2 keV respectively.  
The photoelectric absorption was characterized by 
the {\tt phabs} model.  
Photoionization cross-sections was taken from Balucinska-Church \& McCammon (1992). 
The Galactic 
hydrogen column density was fixed at $N_H=2.3\times10^{20}$ cm$^{-2}$ for Abell~1142, derived 
from the Dickey and Lockman (1990) map incorporated in the {\sl HEASARC} $N_H$ tool.
The particle background was modeled through a number of Gaussian lines and a broken power-law model (see Humphrey et al.\ 2012 and Su et al.\ 2015a).
The background emission of each annulus were forced to have a uniform surface brightness, but the total normalizations were allowed to vary freely.

We used a single thermal {\tt apec} component to model the ICM emission in each annulus.
The ICM components were allowed to vary independently for each radial annulus. 
Since the metallicity is less well constrained than the temperature, we found it necessary to tie together the abundances in the annuli 2 and 3. 
Projected temperature and metal abundance profiles of Abell~1142 are shown in the solid line in Figure~\ref{fig:spec} and the results are listed in Table~1.
To assess the impact of the binning, we also performed the spectral analysis with only three radial bins as shown in the dashed lines, which indicates similar results.

We produced a hardness ratio map of Abell~1142 using the contour binning analysis developed by Sanders (2006).
We generated a contour binning image containing 19 regions for the mosaic image in 0.5-2.0 keV with at least 200 net counts in each bin. 
We extracted counts in each region from the mosaic image in 0.5-2.0 keV and 2.0-7.0 keV respectively. 
We subtracted background components (X-ray background and particle background) from each energy band. 
We obtained a hardness ratio map 
using the ratio of counts in 0.5-2.0 keV to 2.0-7.0 keV bands (see Figure~\ref{fig:hard}). {The uncertainties on hardness ratios are of 20\%--30\%.  In the hard band, the background is dominating the cluster outskirts regions near the chip edges, for which we did not obtain reasonable and meaningful hardness ratio measurement and it was left out from the results. We also derived a hardness ratio profile of 8 radial bins ranging from the cluster center out to 4$^{\prime}$ shown in green circles in Figure~\ref{fig:spec} and tabulated in Table~1.} 

To assess the background systematics, we varied the X-ray background and particle backgrounds by 10\% and 5\% respectively for the spectral analyses. We also varied the level of the blank sky background by 5\% for the hardness ratio measurement. The impacts of these variations are listed in Table~1. Overall, the background systematic uncertainties are smaller than the statistical uncertainties of our results.

\subsection{VLA}

Abell~1142 was observed with the Very Large Array (VLA) at 1.4~GHz in B-array on Jan~12, 1992 and in C-array on Jun~7, 1989 at 1.5~GHz. A total bandwidth of 50~MHz was recorded for both observations The total on-source times were 6 and 7~min for the B and C-array observations, respectively. The data were reduced and calibrated with CASA version 4.4 (McMullin et al.\ 2007). The flux-scale was set using the primary calibrator 3C286 taking the Perley-Butler 2010 flux-scale. The flux-scale was bootstrapped from the primary to the secondary calibrators and the gain solutions were applied to the target field. A few rounds of self-calibration were carried out to refine the gain solution of the target field. The data were then combined and jointly deconvolved to produce the final primary beam corrected image of the cluster. Clean boxes were used during the imaging as well as W-projection (Cornwell et al.\ 2008). We used Briggs et al.\ (1995) weighting with a robust factor of $-0.5$, resulting in an image with a resolution of $8.0\arcsec\times7.4\arcsec$. At the cluster position we measure a r.m.s. noise of 0.37mJy~beam$^{-1}$.

The integrated 1.4~GHz flux density for the source from the NVSS (Condon et al.\ 1998), which should recover all flux, is $42.5\pm1.3$~mJy. This corresponds to a 1.4~GHz radio power of $1.2\times10^{23}$~W~Hz$^{-1}$.

\section{\bf Results}

\subsection{\bf Thermal and chemical distribution}

\begin{figure}[h]
   \centering
  \includegraphics[width=0.48\textwidth]{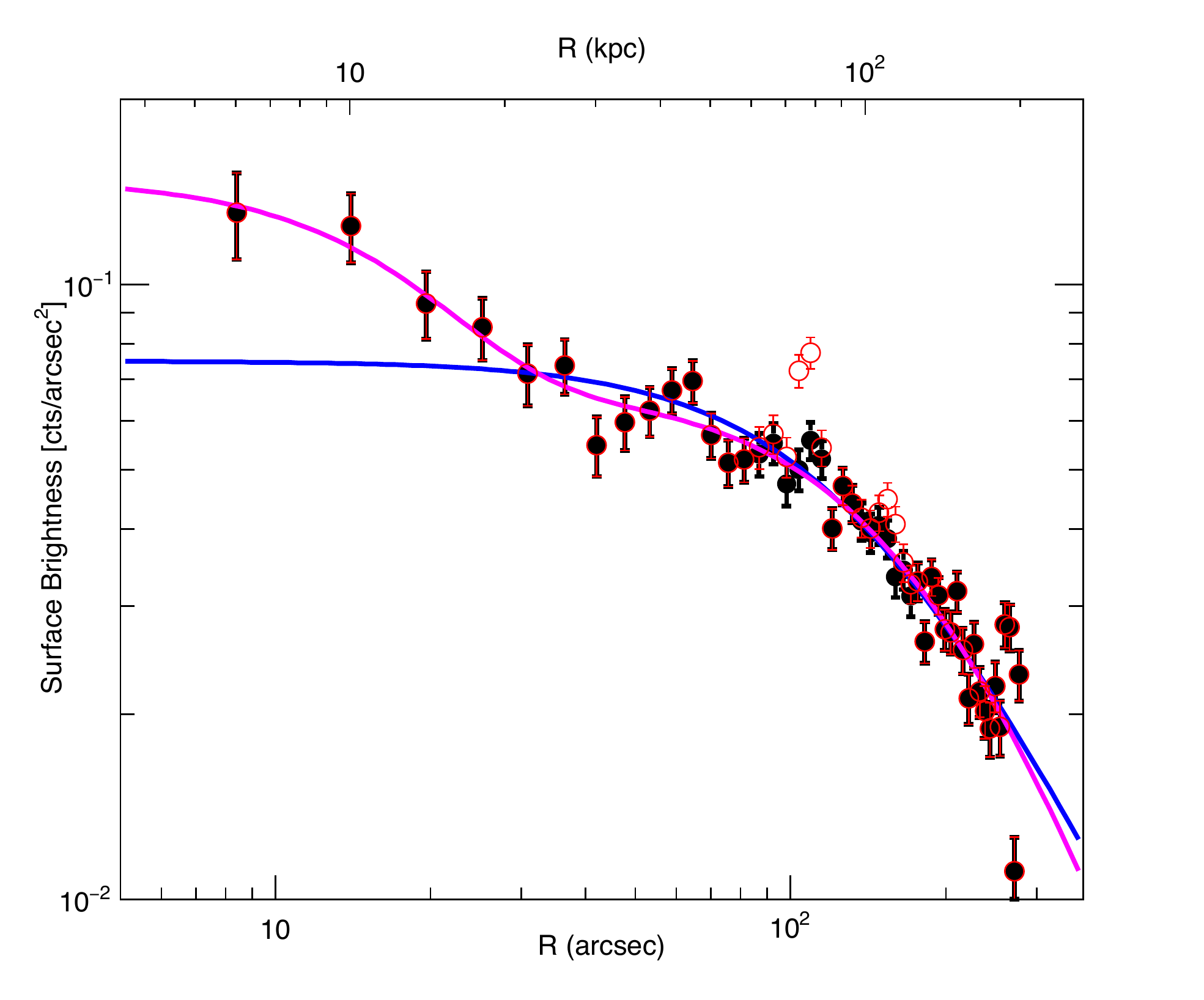}
\figcaption{\label{fig:sur}Surface brightness profile of Abell~1142  in the energy band of 0.5--4.0 keV, centered on the cluster X-ray emission peak (cyan cross in Figure~\ref{fig:A1142}). Blue solid line indicates the best-fit single $\beta$ model with $\beta=0.45\pm0.04$ and $r_c=92\pm16$ kpc. Magenta solid line indicates the best-fit double $\beta$ model with $\beta=0.65\pm0.13$ ($r_{c1}=12\pm4$\,kpc  and $r_{c2}=157\pm31$\,kpc).}
\end{figure}

Despite the overall asymmetric morphology of the X-ray emission, 
we generated the azimuthally averaged surface brightness profile centered on the central X-ray peak as shown in Figure~\ref{fig:sur}. Black (red) circles were produced with the central 20 kpc emissions from Peaks G1 and G2 excluded (included). Emissions from Peaks G1 and G2 are likely to be dominated by the ISM of BCGs. The surface brightness profile with G1 and G2 excluded provides a better trace of the smooth ICM distribution. We fit the profile of black circles to a single $\beta$ model:
$$
S(r)=S_0[1+(\frac{r}{r_c})^2]^{-3\beta+1/2}
$$
We obtained best-fitting parameter values of $\beta=0.45\pm0.04$ and $r_c=92\pm16$\,kpc with a $\chi^{2}/{\rm ndf}$ of $132/47$. 
The fit was significantly improved by applying a double $\beta$ model instead (here we linked the values of the two $\beta$):
$$
S(r)=S_{01}[1+(\frac{r}{r_{c1}})^2]^{-3\beta+1/2}+S_{02}[1+(\frac{r}{r_{c2}})^2]^{-3\beta+1/2},
$$
We obtained best-fits of $\beta=0.65\pm0.13$ and $r_{c1} (r_{c2})=12\pm4\, (157\pm31)$\,kpc with a $\chi^{2}/{\rm ndf}$ of $106/45$.

\begin{figure}[h]
   \centering
    \includegraphics[width=0.53\textwidth]{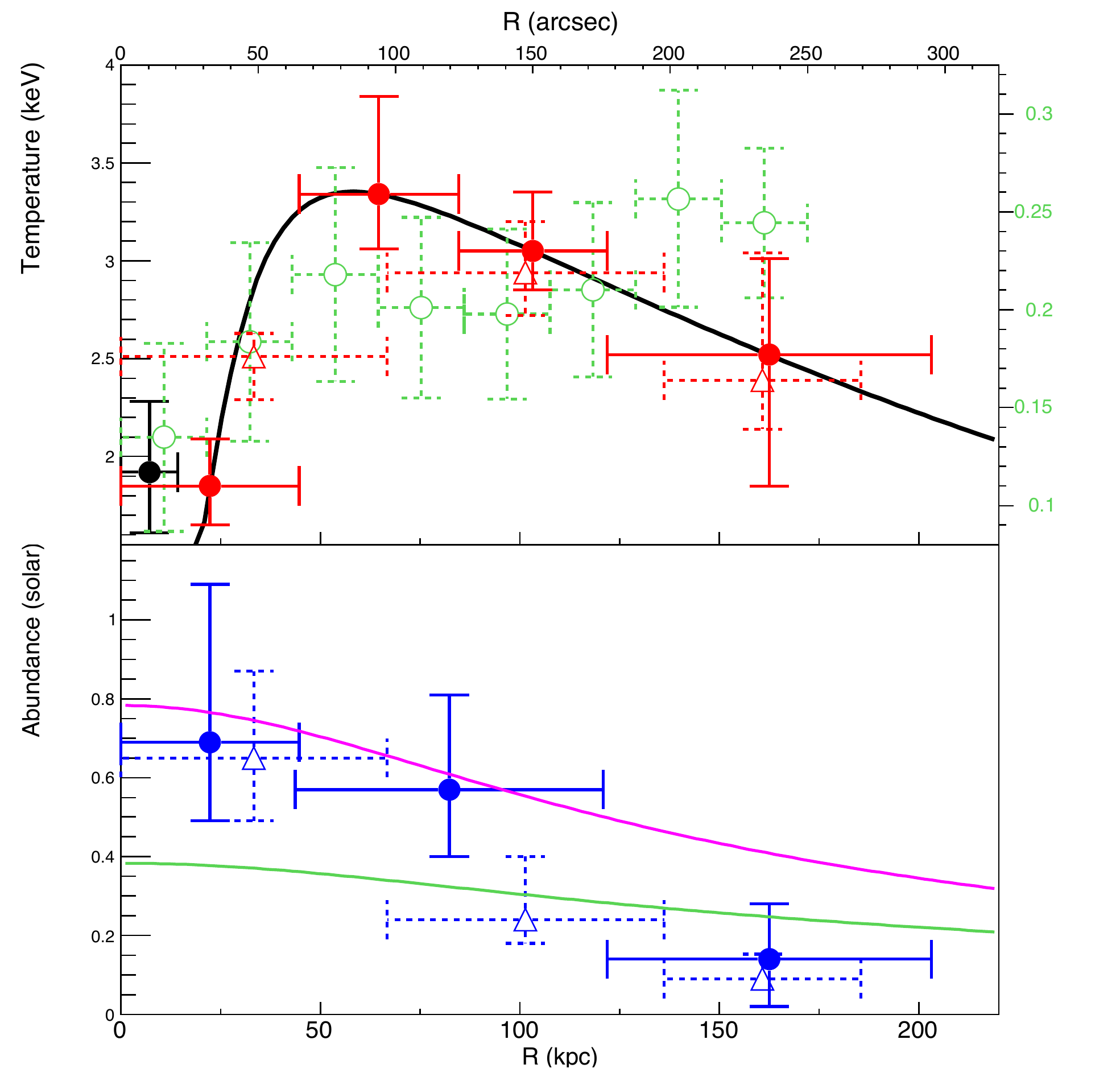}
\figcaption{\label{fig:spec}Projected temperature (top) and metal abundance (bottom) profiles of Abell~1142 centered on the peak of X-ray emission (four magenta annuli in Figure~\ref{fig:hard}). We also show the results of a different radial binning of three annuli (dashed line). Black solid line is the best-fit temperature profile. We added hardness ratio profile in the top panel (green open circles). Magenta (green) solid line is the average metal abundance profile of CC (NCC) clusters studied in Ettori et al.\ (2015).}
\end{figure}

Vikhlinin et al.\ (2006) presented a universal profile to model temperature profiles of cool core clusters.
\begin{equation}
T_{\rm 3D}(r)=\frac{[T_0(r/r_{\rm cool})^{a_{\rm cool}}+T_{\rm min}]}{[(r/r_{\rm cool})^{a_{\rm cool}}+1]}\times\frac{(r/r_t)^{-a}}{[1+(r/r_t)^b]^{c/b}}
\end{equation}
We fit the projected temperature profile of Abell~1142 to this universal profile and obtained a best-fit of [$T_0, r_{\rm cool}, a_{\rm cool}, T_{\rm min}, r_t, a, b, c$] = [$3.93, 23.9, 3.13, 0.044, 380.7, 0.012, 1.36, 2.24$] as shown in Figure~\ref{fig:spec}. {These parameters are poorly constrained.}

\begin{figure}[h]
   \centering
    \includegraphics[width=0.5\textwidth]{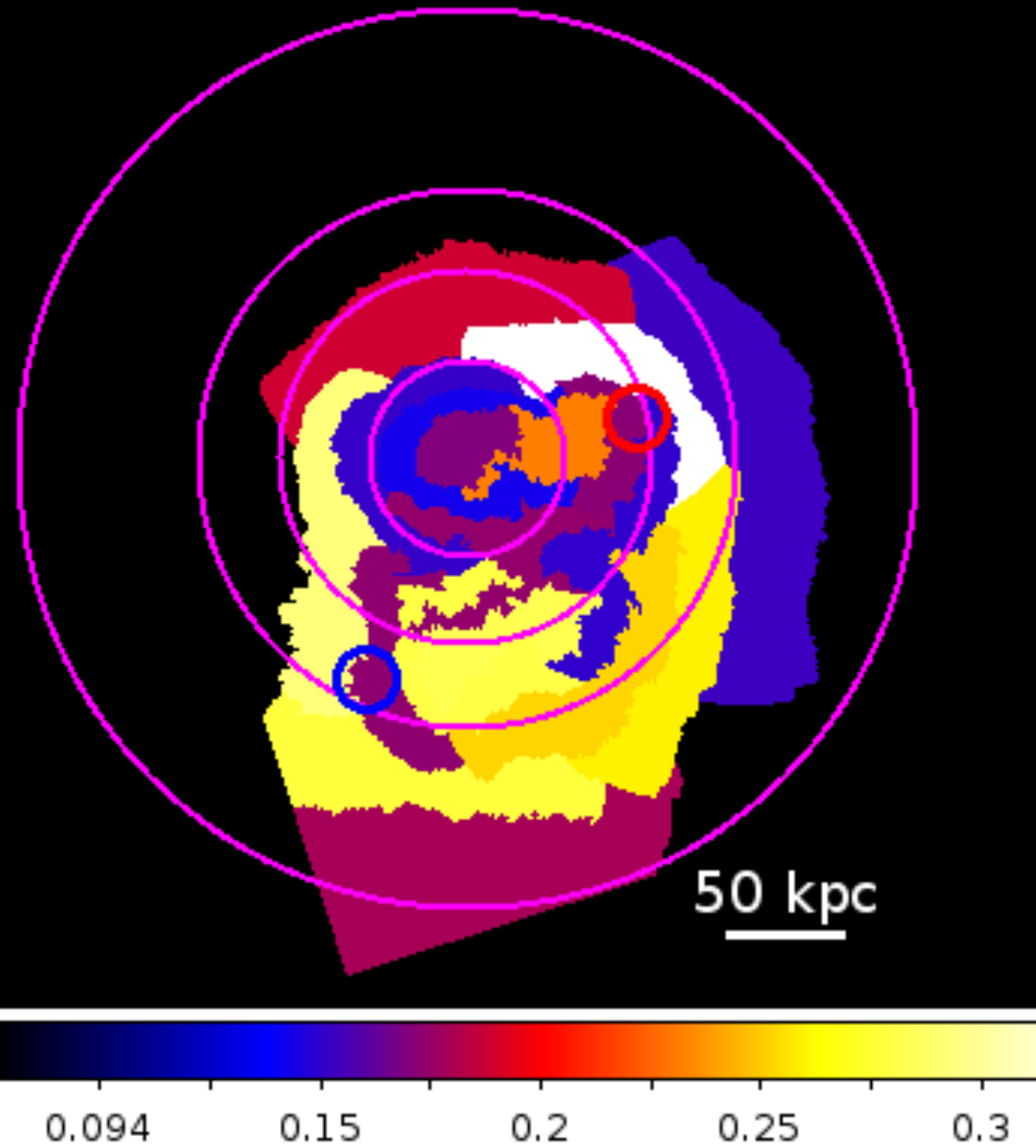}
\figcaption{\label{fig:hard}Contour binning hardness ratio map  of Abell~1142 (2.0--7.0 keV to 0.5--2.0 keV). Blue and red circles indicate the locations of two brightest cluster galaxies. Magenta annuli, centered on the peak of X-ray emission, represent extraction regions of spectral analysis (Figure~\ref{fig:spec}).}
\end{figure}

Abell~1142 resembles a scaled-down version of the Coma Cluster which is a massive NCC cluster containing two dominant early-type galaxies (NGC~4874 and NGC~4889) with the cluster X-ray emission peak residing between them. 
However, unlike the Coma Cluster, the X-ray peak of Abell~1142 is substantially cooler than the surroundings, as presented in its Figure~\ref{fig:spec} and Figure~\ref{fig:hard}. 
We observe a temperature below 2.0 keV towards the center and it rises to $3.5$ keV just outside 50\,kpc then declines to 2.5 keV out to 200 kpc. In addition, its metal abundance profile reaches $0.7$\,Z$_{\odot}$ at the center and gradually declines to $\sim$0.1\,Z$_{\odot}$ out to 200 kpc.
The behaviors of the temperature and metal abundance profiles are very typical of those observed in cool-core clusters (De Grandi \& Molendi 2001; Maughan et al.\ 2012). 
{We also fit the spectrum extracted from the central 20$^{\prime\prime}$ but fixed its metallicity at the current best-fit value of the central 65$^{\prime\prime}$ and obtained a best-fit temperature of $1.92^{+0.36}_{-0.31}$ keV and an electron density $n_e$ of $5.8\pm0.3\times10^{-3}$\,cm$^{-3}$, where $n_e$ is calculated through $$
{\rm norm} = \frac{10^{-14}}{4\pi[D_A(1+z)]^2}\int{n_en_H}{dV}. 
$$}

 \begin{deluxetable*}{c|ccccccccc}

\tablewidth{0pc}
 \centering
\tablecaption{Best fits and background systematics}
\startdata

\hline
&& \multicolumn{2}{c}{annulus 1}& \multicolumn{2}{c}{annulus 2}& \multicolumn{2}{c}{annulus 3} & \multicolumn{2}{c}{annulus 4}\\
\hline
 \multirow{2}{*}{Temperature}&best-fit& \multicolumn{2}{c}{$1.85^{+0.24}_{-0.20}$}& \multicolumn{2}{c}{$3.34^{+0.5}_{-0.28}$} &\multicolumn{2}{c}{$3.05^{+0.3}_{-0.2}$} &\multicolumn{2}{c}{$2.52^{+0.49}_{-0.67}$}\\
&{$\Delta$ Particle}&\multicolumn{2}{c}{-0.03 (-0.01)} &\multicolumn{2}{c}{ -0.03 (+0.01)}&\multicolumn{2}{c}{-0.05 (+0.01)}&\multicolumn{2}{c}{-0.04 (+0.07)}\\
(keV)&{$\Delta$ X-ray}& \multicolumn{2}{c}{-0.01 (-0.01)}& \multicolumn{2}{c}{+0.01 (+0.03)}&\multicolumn{2}{c}{-0.01 (+0.01)}&\multicolumn{2}{c}{-0.01 (+0.04)}\\
\hline
 \multirow{2}{*}{Metallicity}& best-fit&\multicolumn{2}{c}{ $0.67^{+0.4}_{-0.2}$}&\multicolumn{2}{c}{$0.57^{+0.24}_{-0.17}$} &\multicolumn{2}{c}{~~ -} &\multicolumn{2}{c}{$0.14^{+0.14}_{-0.12}$}\\
&$\Delta$ Particle  & \multicolumn{2}{c}{-0.01 (+0.01)}&\multicolumn{2}{c}{ -0.03 (+0.04)}&\multicolumn{2}{c}{~~  -}&\multicolumn{2}{c}{+0.02 (+0.01)}\\
(Z$_{\odot}$)&$\Delta$X-ray& \multicolumn{2}{c}{+0.03 (-0.02)}&\multicolumn{2}{c}{ +0.01 (-0.02)}&\multicolumn{2}{c}{~~ -}&\multicolumn{2}{c}{+0.01 (-0.02)}\\
\hline
&&ann 1 &ann 2 &ann 3 &ann 4 &ann 5 &ann 6 &ann 7 &ann 8\\
\hline

 {Hardness}&best-fit& $0.14\pm0.05$&$0.18\pm0.05$&$0.22\pm0.06$ &$0.2\pm0.05$&$0.2\pm0.04$ &$0.21\pm,0.05$ &$0.26\pm0.06$&$0.24\pm0.04$\\
 ratio&$\Delta$background&$\pm$0.02 &$\pm$0.02 &$\pm$0.03 &$\pm$0.03 &$\pm$0.03&$\pm$0.04 &$\pm$0.05 &$\pm$0.05
\enddata
\tablecomments{We increase (decrease) best-determined particle and X-ray backgrounds by 5\% and 10\% respectively for the spectral analyses. We varied the level of blank sky by 5\% for the hardness ratio measurements. The background systematic uncertainties are smaller than statical uncertainties. Best fits are plotted in Figure~3. }
\end{deluxetable*}

\subsection{\bf Galaxies in Abell~1142}

\begin{figure}[h]
   \centering
    \includegraphics[width=0.5\textwidth]{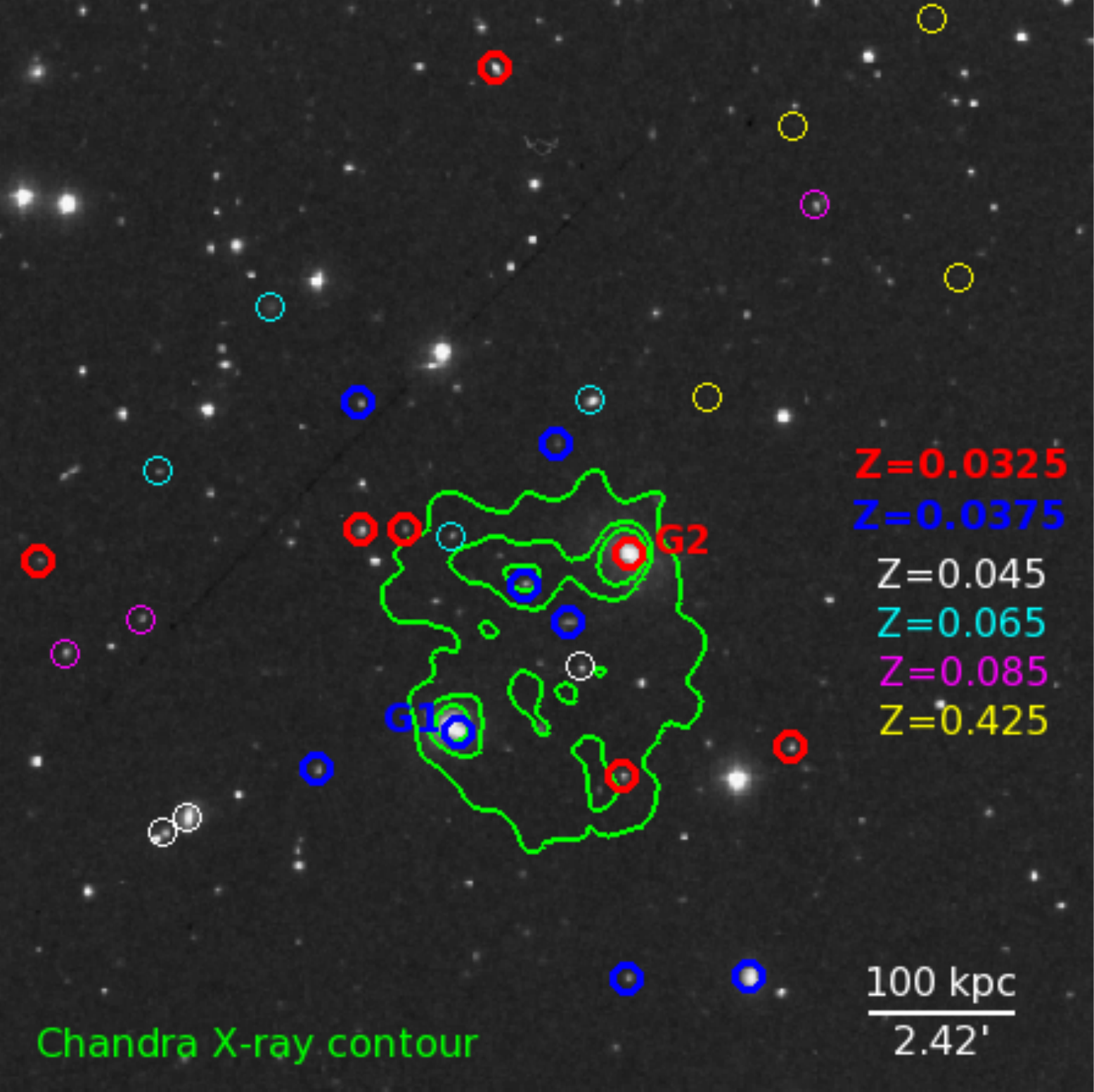}
\figcaption{\label{fig:dss}. DSS image galaxies of the central 18\arcmin$\times$18\arcmin of Abell~1142. {\sl Chandra} X-ray contour is based on a adaptively smoothed image; contour levels are at [0.9, 1.5, 2.1]$\times 10^{-9}$ photon\,cm$^{-2}$\,s$^{-1}$. We overlaid the redshift distributions of SDSS galaxies (only those with at least three galaxies per redshift bin are shown here).}
\end{figure}

\begin{figure}[t]
\centerline
{\includegraphics[height=0.35\textheight,width=0.5\textwidth]{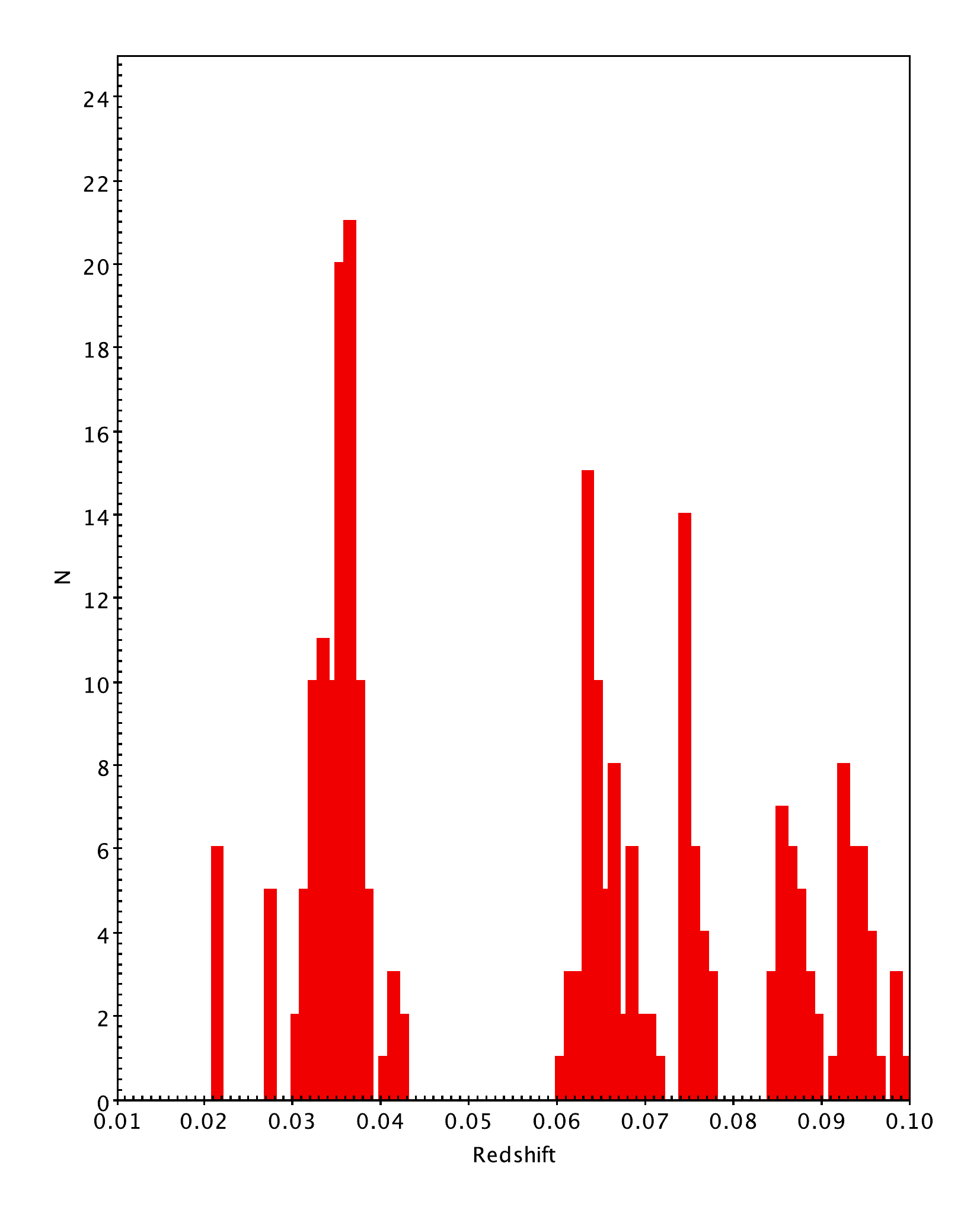}}
\figcaption{\label{fig:redshift2} Histogram of all the available 
redshifts in the SDSS database and NED within a 60\arcmin\ field (galaxies with $z > 0.1$ are not shown for display purposes). 
The Abell~1142 group stands out as the redshift peak at $z=0.035$. 
 }
\end{figure}

We selected all galaxies with known recession velocities from NED and SDSS within 60\arcmin\ (2.4 Mpc at the redshift of the source). The SDSS Data are augmented by the redshift surveys of the object performed by Geller et al.\ (1984), Mahdavi et al.\ (2004) and Quintana et al.\ (1995).
Group members have been selected through the elimination of background and foreground 
galaxies along the line of sight to Abell~1142. As a first step we applied 
the ``velocity gap" method (De-Propris et al.\ 2002) sorting the galaxies in redshift space and 
calculating their velocity gap, defined for the $n$th galaxy as 
$\Delta v_n\;=\;cz_{n+1}-cz_{n}$.
We uses 1000 km\,s$^{-1}$ as velocity gap. Figure~\ref{fig:redshift2} shows that Abell~1142 
is clearly detected as a peak at $z=0.03-0.04$ populated by 102 candidate 
member galaxies. In Figure~\ref{fig:dss}, we marked out the locations of galaxies with accurate SDSS redshifts on the DSS optical image in the central 18\arcmin$\times$18\arcmin. The distribution of galaxies in the range of $z$ = 0.03--0.04 overlaps with the X-ray emission contour of the {\sl Chandra} observation of Abell~1142. In contrast, the distribution of galaxies in other redshift bins appears to be offset from the X-ray emission in projection. The X-ray emission, in particular the emission peak, is unlikely to be a background cluster.

As a second step we used the ``shifting gapper" method (Fadda et al.\ 1996). 
In this method interlopers are separated from members by exploiting both 
radial and peculiar velocity information. The data are binned radially such that
each bin contains at least 20 objects. The peculiar velocities of the galaxies are
calculated as $v_{\rm{pec}} = c((1 + z_{\rm{pec}})^2 - 1)/((1 + z_{\rm{pec}})^2 + 1)$, 
where $c$ is the speed of light, $z_{\rm{pec}} = (z_{\rm{gal}} - z_{\rm{cos}})/(1 + z_{\rm{cos}})$
is the peculiar redshift of the galaxy and $z_{\rm{cos}}$ is the mean group velocity
estimated by the biweight location estimator (Beers et al.\ 1990) and assumed
to represent the cosmological redshift of the group.
Within each bin galaxies are sorted by peculiar velocity with velocity gaps determined in peculiar 
velocity.
Galaxies are rejected as interlopers if their peculiar velocity gaps are above
the value of the ``f pseudosigma" (Beers et al.\ 1990), an estimator robust
to the presence of outliers. The above procedure was iterated until the number of members was stable.

\begin{figure}[h]
\centerline
{\includegraphics[height=0.53\textheight]{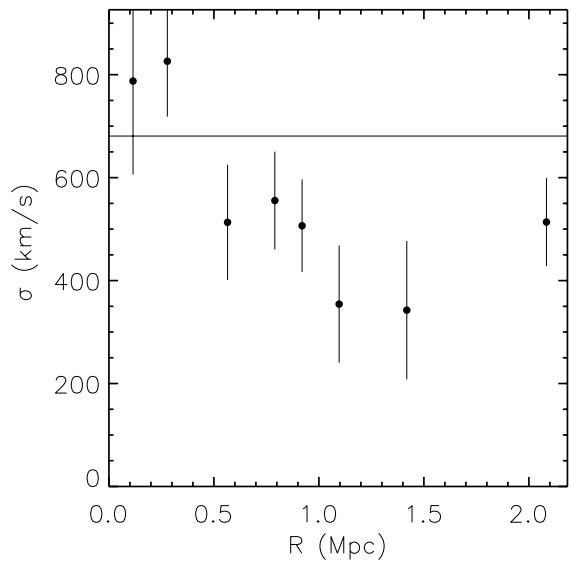}}
 \vspace{-4cm}
\figcaption{\label{fig:disp}  
Velocity dispersion profile of the Abell~1142 group, where the center of the group is 
assumed to be at the position of the galaxy IC 664. 
Bins contain 10 galaxies each and 1$\sigma$ jackknife errors are shown.
The solid line show the expected value of dispersion velocity assuming 
density-energy equipartition between IGM and galaxies, i.e. $\beta_{spec}=1$}
\end{figure}

The shifting gapper is able to exclude a group of high velocity galaxies 
($cz_{\rm{gal}} > 12500$ \,km\,s$^{-1}$) and a group of low velocity galaxies 
($cz_{\rm{gal}} < 9500$ \,km\,s$^{-1}$) already discussed by Geller et al.\ (1984) as interlopers
associated to the nearby galaxy cluster Z1056.9+0922.
The final group sample contains 88 members. The value for the biweight location 
estimator of the mean group velocity is $10614\pm73$ km\,s$^{-1}$ which corresponds to 
$z_{\rm{cos}}=0.0354\pm0.0002$.
We used the biweight scale estimator to estimate a velocity dispersion of 
$575\pm36$ km\,s$^{-1}$.
The errors for the redshift and velocity dispersion are 1$\sigma$ and are estimated 
using the jackknife resampling technique.

\begin{figure}[t]
\centering
\includegraphics[height=0.3\textheight,width=0.55\textwidth]{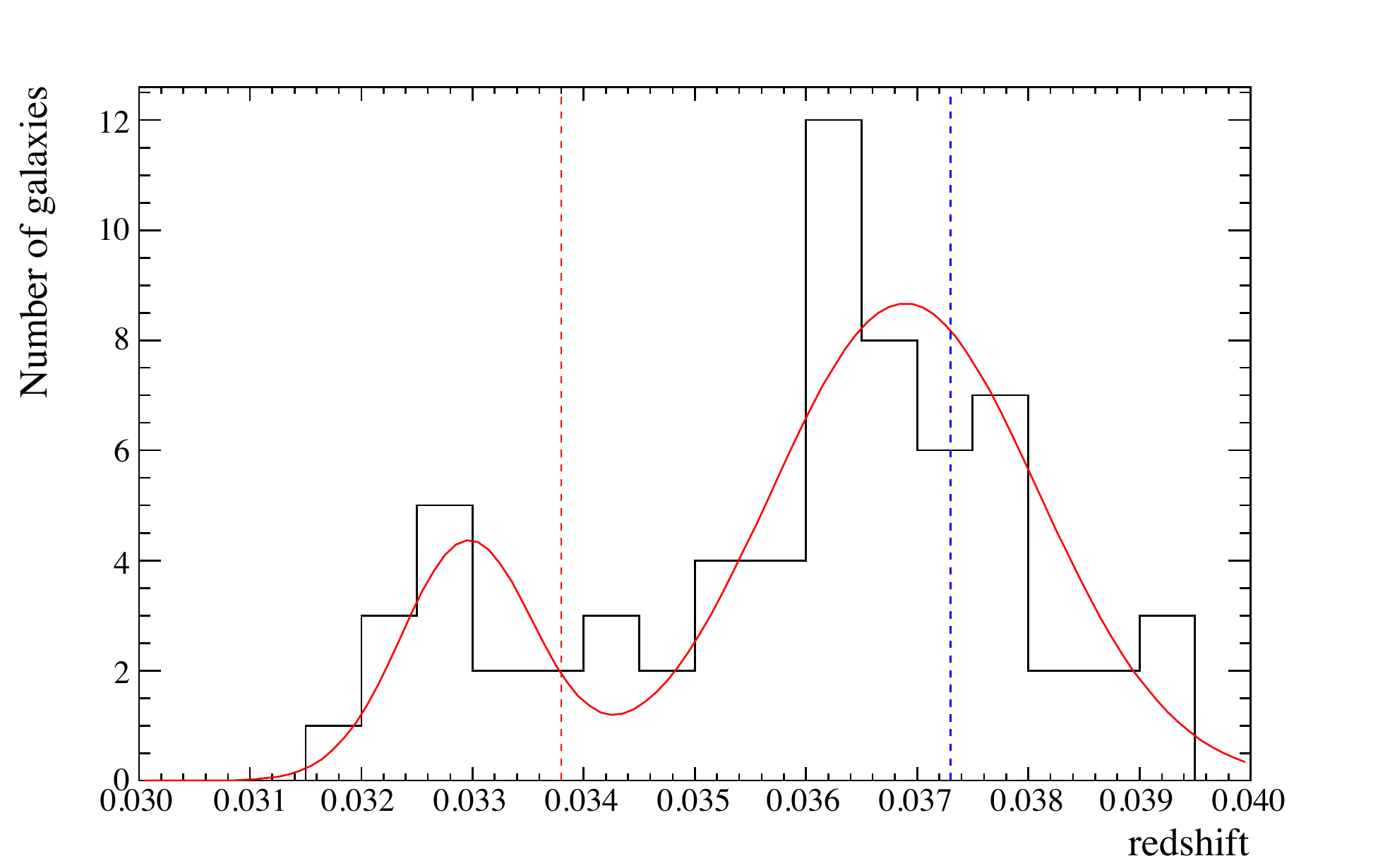}
\figcaption{\label{fig:histo}Histogram of the redshifts of the final
sample of members of A1142 within 1.5 Mpc using a bin size of $\Delta z=0.0005$. Blue and red dashed lines indicate the redshifts of G1 and G2 respectively.}
\end{figure}

In Figure~\ref{fig:disp} we show the velocity dispersion profile derived 
for the data. 
Bins contain 10 galaxies each and velocity dispersions have been calculated using 
the gapper algorithm (Beers et al.\ 1990).
The velocity dispersion profile shows some interesting features: it flattens
at high radii possibly due to a robust asymptotic value in the external 
group regions, as found for the majority of nearby clusters (e.g., Fadda et al.\ 1996, Girardi et al.\ 1996). The increase at radii larger than 2 Mpc indicates a likely interlopers contamination. 
Furthermore the relation used to calculate $\rm{R_{200}}\footnote{The radius within which the enclosed average density is 200 times the critical density of the Universe.} = \sqrt{3}\sigma/10H(z)$ (Carlberg et al.\ 1997) gives a value of 1.4 Mpc. 
We therefore restrict the member galaxies to the 67 ones within 1.5 Mpc. Their redshift distribution
shows a bimodal distribution as shown in Figure~\ref{fig:histo}. 
We fit this histogram to a model consisting of two gaussian distributions. Each distribution peaks at  at $z=0.033$ and $z=0.037$ respectively. This corresponds to a 1200 km\,s$^{-1}$ rest frame velocity difference with velocity dispersions of 175 km\,s$^{-1}$ and 360 km\,s$^{-1}$.
We marked these two populations in blue and red respectively in Figure~\ref{fig:dss}. They overlap in projection with G1 and G2 each belonging to one population. 
The radial velocity difference between G1 ($z_{\rm G1}=0.0373$) and G2 ($z_{\rm G2}=0.0338$) is only 1000 km\,s$^{-1}$.

We can therefore conclude that there is evidence for a merger in Abell~1142.
In particular the high dispersion velocity in the core ($\sim 800$ \,km\,s$^{-1}$) is indicative of the
presence of substructures.
We speculate that a subgroup comprising G2 is merging with the main body of Abell~1142 to which G1 belongs. A likely lower
limit on the velocity of the merger can be estimated by the line of sight difference of 
velocities between the mean of the components of G1 and G2, therefore 1000 \,km\,s$^{-1}$. 
This merger is mostly at the cluster center while the larger radii of the cluster is relatively relaxed.

 \subsection{\bf The BCGs}
 
We note that G1 is made up of three individual galaxies, instead of a single galaxy, as shown in the SDSS $R$-band image in Figure~\ref{fig:dss}, labeled as Objects 1, 2, and 3 (Object 0 is a foreground star). All of them appear to be elliptical galaxies. SDSS measured a spectroscopic redshift of Object 1 of $z=0.03729$, consistent with the average of ``the G1 subcluster". The spectroscopic redshifts of Object 2 and 3 can not be obtained through SDSS since they are separated by $\lesssim 3^{\prime\prime}$ from each other, falling short of the fiber size of SDSS. 
The redshifts of Objects 2 and 3 found in literature are similar to Object 1 but their values and positions are inconsistent. Reliable redshift measurements are currently unavailable, although they seem unlikely to be background objects.

We also show the {\sl Chandra} X-ray image, {\sl VLA} radio image, and {\sl 2MASS} $K$-band images of the same field-of-view in Figure~\ref{fig:dss}. Objects 1 and 2 contain X-ray nuclei detected by {\tt wavedetect}. We inspected the 2.0--8.0 keV image and confirmed that they are point sources. They also contain point sources in the radio band. Objects 1 and 2 may each harbor an AGN. We obtained count rates from point source regions given by {\tt wavedetect}. We derived X-ray luminosities of 1.6$\pm$0.7\,$\times10^{40}$ erg\,s$^{-1}$ and 6.3$\pm$1.1\,$\times10^{40}$ erg\,s$^{-1}$ in the 2--8 keV band for the AGN in Objects 1 and 2 respectively assuming a powerlaw spectrum with an index of 1.4. All these three galaxies appear to have similar $R$-band and $K$-band luminosities. They may have similar stellar masses. 

Member galaxies orbiting in the ICM would gradually lose angular momentum to the background ICM, stars, or dark matter through dynamical friction.
This effect becomes progressively significant towards cluster center (El-Zant et al.\ 2004; Gu et al.\ 2013). Given their current positions (possibly within a radius of 5 kpc), these three galaxies are anticipated to sink into the center of G1 subcluster within a time much shorter than the infall time in a cuspy dark matter halo (Cole et al.\ 2012). 
{While the projected distance of the galaxies in question is unknown presently, if they all lie within $\sim$5kpc of the BCG it would have important consequences for the shape of the dark matter profile.} The four galaxies found in the 10 kpc core of Abell~3287 is another example (Massey et al.\ 2015).

 We derived the $K$-band infrared luminosities of G1 and G2 using {\sl 2MASS} archived images (Skrutskie et al.\ 2006). 
We obtained an average $L_{\rm K, G1}=2\times10^{11}$L$_{\rm K,\odot}$ for each contained in G1 (one third of the total $K$-band luminosity of G1) and obtained $L_{\rm K, G2}=2.66\times10^{11}$L$_{\rm K,\odot}$ for G2, IC~664.
 
 \begin{figure*}[h]
   \centering
    \includegraphics[width=0.9\textwidth]{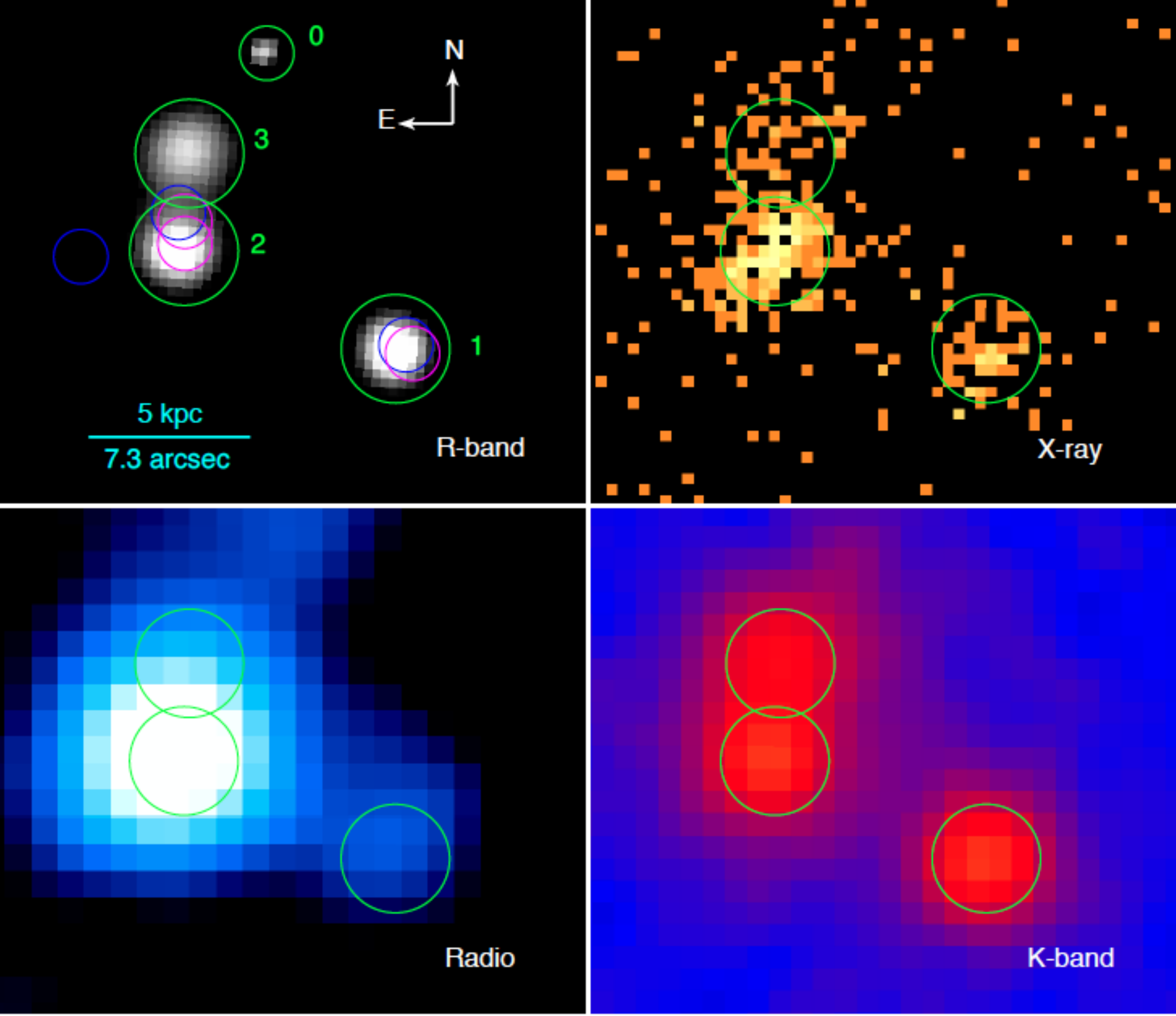}
    
    \figcaption{\label{fig:3492}. $R$-band (SDSS), X-ray ({\sl Chandra}), Radio ({\sl VLA}), and $K$-band (2MASS) images of G1 (NGC~3492) on the same scale. Objects 1, 2, 3 are early-type galaxies. Object 0 is a foreground star. Magenta circles are positions of three galaxies listed in NED and blue circles are three galaxies listed in HyperLeda.}
\vspace{0.25cm}
 \includegraphics[width=0.85\textwidth]{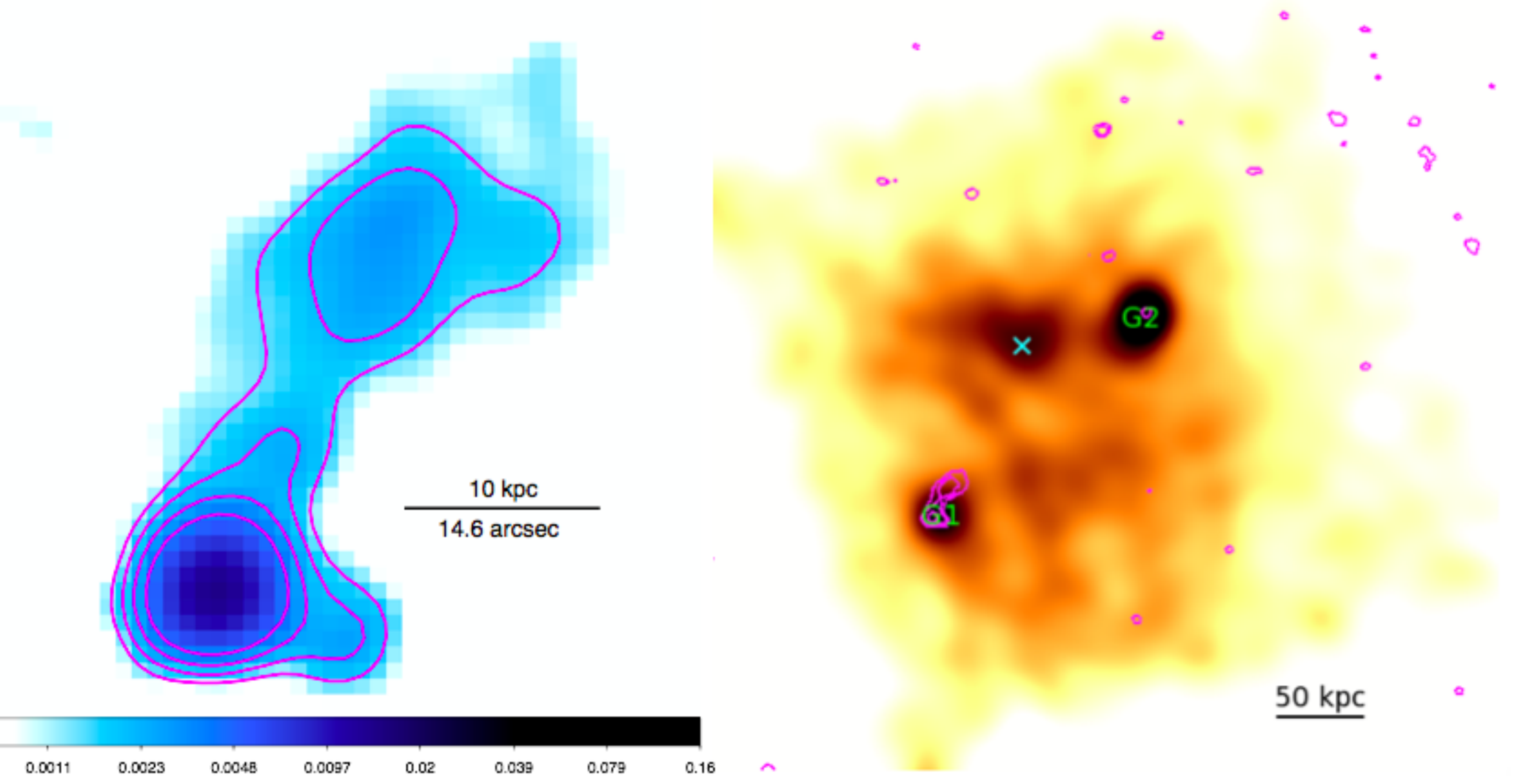}
\figcaption{\label{fig:radio}. {\it left}: {\sl VLA} radio image of G1 showing extended radio emission in the northwest direction in the unit of Jy beam$^{-1}$. Magenta contour levels are set at sqrt[1, 2, 4, 8]$\times$\,4$\sigma_{\rm rms}$ where $\sigma_{\rm rms}$=0.37 mJy beam$^{-1}. ${\it right}: {\sl Chandra} X-ray image of Abell~1142 with radio contour overlaid.}
\end{figure*}

\begin{figure}[h]
   \centering
    \includegraphics[width=0.5\textwidth]{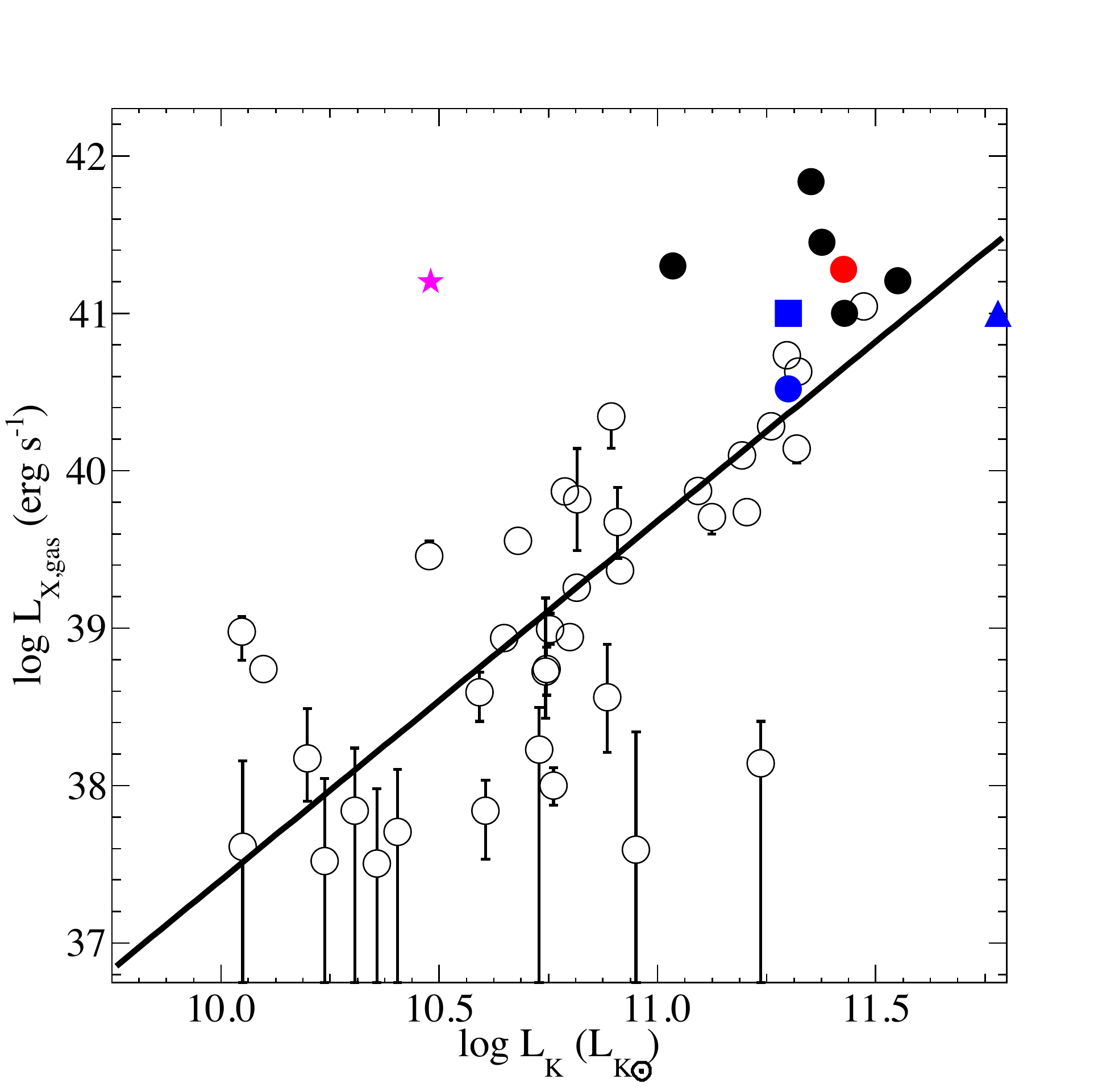}
\figcaption{\label{fig:lxk} Gaseous X-ray luminosities as a function of $K$-band luminosities for all galaxies taken from Su et al.\ (2015b). The black solid line is the best-fit $L_{X_{\rm gas}}$--$L_K$ relation. Black solid circles indicate group center galaxies. Red circle represents IC664 (G2). Blue symbols represent NGC~3492 (G1)-- blue circle: using the average $L_{X_{\rm gas}}$ and $L_K$ of the three galaxies in G1; blue square: assuming all the $L_{X_{\rm gas}}$ is from one of the three galaxies; blue triangle: treating all three galaxies as one galaxy (total  $L_{X_{\rm gas}}$ and $L_K$). Magenta star represents the small galaxy at the central X-ray peak.}
\end{figure}

We were unable to perform X-ray spectral analysis of the faint ISM of G1 or G2 due to the limited exposure. Based on their net count rates, we estimated gaseous X-ray luminosities within two effective radii of G1 and G2 using PIMMS, assuming an ISM temperature of 0.5 keV and a metal abundance of 0.7 Z$_{\odot}$ (typical of early-type galaxies, Su \& Irwin 2013; Su et al.\ 2015b). 
We excluded contributions from Low Mass X-ray Binaries and stellar diffuse emission based on their $K$-band luminosities\footnote{For stellar diffuse emission (cataclysmic variables and active binaries), we use the relation calibrated by Revnivtsev et al.\ (2008): $L_{X_{\rm CV/AB}}/L_K= 5.9\times10^{27}$\,erg\,s$^{-1}$ ${L_{K\odot}}^{-1}$. For Low Mass X-ray Binaries, we use the relation calibrated by Boroson et al.\ (2011): $L_{X_{\rm LMXB}}/L_K= 10^{29}$\,erg\,s$^{-1}$ ${L_{K\odot}}^{-1}$.}. This leads to a gaseous X-ray luminosity to $K$-band luminosity ratios of $5\times10^{28}$\,erg\,s$^{-1}$ ${L_{K\odot}}^{-1}$ and $7\times10^{29}$\,erg\,s$^{-1}$ ${L_{K\odot}}^{-1}$ for G1 and G2 respectively (We assume all three G1 galaxies have the same diffuse X-ray luminosities). We compared G1 and G2 with other early-type galaxies studied in Su et al.\ (2015b) in Figure~\ref{fig:lxk}.
Group center galaxies are marked in black solid circles. 
The $L_{\rm X, gas}/L_{\rm K}$ of G1 and G2 are typical of group center galaxies. This result reinforces the status of G1 and G2 as BCGs.

The {\sl VLA} image of G1 reveals that it contains a radio tail pointing northwest, approximately 10 kpc wide and 30 kpc long (Figure~\ref{fig:radio}). It appears to be more extended than other background radio sources found in the same {\sl VLA} observation. It may be a radio lobe ejected by Object 2.
The direction of this tail suggests that G1 may move southeastward in the plane of the sky. 
We note that while G1 harbors a radio lobe G2 only contains a small radio point source. This is consistent with our speculation that G2 is the center of a subcluster merging with the main body of Abell~1142 centered at G1.

We note that there is a galaxy located right at Peak X with nearly the same redshift as G1 (Figure~\ref{fig:dss}). 
Its $K$-band luminosity is only $3\times10^{10}$\,$L_{K\odot}$, 10 times smaller than G1 and G2. 
Its X-ray emission associated is very unlikely to be dominated by the ISM of this relatively small galaxy. First, the temperature of the gas inferred in the smallest aperture is close to 2 keV which implies a group-scale, rather than a galaxy-scale, halo. Second, the $L_{\rm X, gas}/L_{\rm K}$ ratio would place this object far off the relation in Figure~\ref{fig:lxk} (magenta star). 
For these reasons, it is most likely that the X-ray peak is dominated by cluster emission.


\section {\bf Discussion}

Abell~1142 is a unique object invoking some important implications discussed in the following subsections.

\subsection{\bf Is Abell~1142 a cool-core cluster?} 

Several lines of evidences suggest that Abell~1142 is a CC cluster: the positive temperature gradient, a negative metal abundance gradient, and the need of a double $\beta$ model, although its morphology and the offset between its X-ray emission peak and  BCG are not so typical of CC clusters. 

Quantitatively distinguishing CC and NCC clusters is more problematic than it might seem at first, because even classic morphologically disturbed NCC clusters like Coma can possess isolated regions of temperature drops. Here we consider two criteria that have been proposed to distinguish a CC from a NCC cluster.
One criteria is that the ICM cooling time at the center of CC (NCC) clusters should be shorter (longer) than the look back time to $z=1$ (i.e., 7.7 Gyr) (Birzan et al.\ 2004; Hudson et al.\ 2010). The other criteria is that CC (NCC) clusters should have a central entropy smaller (larger) than 70 keV cm$^2$ (Mahdavi et al.\ 2013).
{Using the best-fit electron density and temperature determined for the central 20$^{\prime\prime}$ (14 kpc), we calculate the cooling time profile from Voigt \& Fabian (2004) as
$$
t_{\rm cool}=20 \left(\frac{n_e}{10^{-3} {\rm cm}^{-3}}\right)^{-1}\left(\frac{T}{10^{7} {\rm K}}\right)^{0.5} {\rm Gyr}
$$
and the entropy profile $S = {\rm k}T/{n_e}^{2/3}$. We obtained $t_{\rm cool}=5.1^{+0.8}_{-0.7}$ Gyr and $S=59.6^{+13.7}_{-11.3}$\,keV\,cm$^2$, marginally satisfying the criterion of cool core. } 

The difference between CC and NCC clusters is not only in thermo-dynamics but also in chemo-dynamics. The central gas in CC clusters is metal richer than NCC systems. 
Using {\sl XMM-Newton}, Ettori et al.\ (2015) studied the metal abundance distribution of 83 clusters out to a radius of $>0.4\,R_{500}$. Their work provided a representative radial and redshift dependence in the form of $Z(r,z) = Z_0 (1+(r/0.15R_{500})^2)^{-\beta} (1+z)^{-\gamma}$, with $(Z_0, \beta, \gamma) = (0.83\pm0.13, 0.55\pm0.07, 1.7\pm0.6)$ for CC clusters and $(Z_0, \beta, \gamma) = (0.39\pm0.04, 0.37\pm0.15, 0.5\pm0.5)$ for NCC clusters. We compared the metal abundance profile of Abell~1142 to these average profiles of CC and NCC clusters in Figure~\ref{fig:spec} (bottom). The central metallicity of Abell~1142 is very typical of CC clusters.

Overall, the current results suggest that it is entirely possible for Abell~1142 to harbor a cool core at its center although not very pronounced. Note that these values are very uncertain. Our using of projected profiles to derive these parameters may also smear out ``cool core" features.
In addition, these standards are very arbitrary and may not apply well to a lower-temperature and asymmetric system like Abell~1142. On the other hand, if the temperature of the surrounding hot gas was greatly enhanced due to shock heating, the relative significance of this ``cool core" would be undermined.

\subsection{\bf Ongoing merging scenario}

We estimate a velocity dispersion of Abell~1142 of 575 km\,s$^{-1}$.
 This corresponds to a thermal temperature of 2.6 based on the $\sigma-T_{X}$ relation calibrated by Wu et al.\ (1999) for galaxy clusters.  Bohringer et al.\ (2000) listed a {\sl ROSAT} X-ray luminosity of $L_{\rm X}=2.8\times10^{43}$erg\,s$^{-1}$ in 0.1--2.4 keV for Abell~1142, which translates to a $T_{X}$ of 1.4\,keV based on the $L_{X}-T_{X}$ relation of Wu et al.\ (1999). 
 Note that there are large internal scatters of the $\sigma-T_{X}$ and $L_{X}-T_{X}$ relations as well. 
Nevertheless, the fact we observe a $T_{\rm X}$\,=\,3.5 keV of Abell~1142 is higher than we expect and could be a sign of recent shock heating caused by a merger.

\begin{figure}[h]
   \centering
   \hspace{-1.2cm}
    \includegraphics[width=0.55\textwidth]{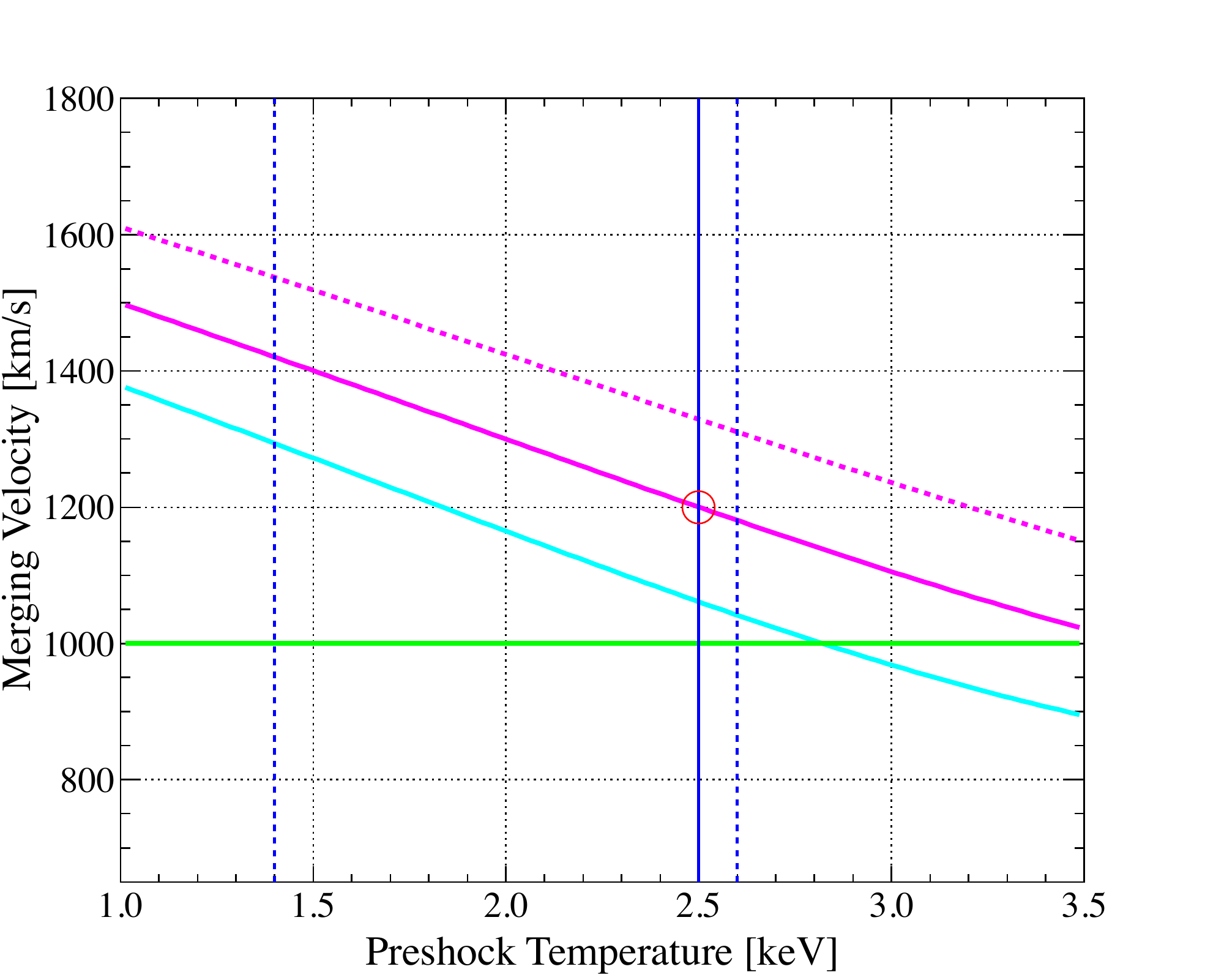}
\figcaption{\label{fig:vt} Estimated merging velocities between the two subclusters as a function of the ICM preshock temperatures for Abell~1142. Cyan and magenta solid (dashed) lines represent postshock temperatures of 3.5\,keV and 4.0\,keV (4.5\,keV) respectively. Green solid lines indicates 1000 km\,s$^{-1}$, the radial merging velocities inferred from the redshift distributions of member galaxies. Two blue dashed lines indicate 1.4\,keV and 2.6\,keV covering the range of preshock temperatures inferred from the cluster $L_{\rm X}$ and $\sigma$ respectively. We marked a preshock temperature of 2.5 keV in blue line, inferred from the temperature profile of Abell~1142 out to 200 kpc. Red circle indicates the most likely scenario.}
\end{figure}

Abell~1142 may contain two subclusters, likely with G1 and G2 being their BCGs. These two subclusters may be experiencing merging with a velocity difference of  $1000$\,km\,s$^{-1}$ in the line of sight. The thermal temperature (preshock temperature) of Abell~1142 should be in the range of 1.4-- 2.6 keV. The fact we observe a $\sim$ 3.5\,keV hot gas indicates the effect of shock heating caused by this merger.
The temperature jump is directly related to the Mach number through the Rankine-Hugeniot equation
$$
\frac{T_2}{T_1}=\left[\frac{2\gamma\mathcal{M}^2-(\gamma-1)}{(\gamma+1)^2}\right]\left[\gamma-1+\frac{2}{\mathcal{M}^2}\right].
$$
The Mach number is expected to be the ratio of the merging velocity to the sound speed ($\mathcal{M}=v/c_s$) and the sound speed in the plasma can be expressed as a function of the preshock temperature ($T_1$) as $c_s=\sqrt{(\gamma{\rm k}T_1)/(\mu m_{\rm H})}$, where $\gamma=5/3$ and $\mu=0.62$. 
These connections allow us to express the merging velocity as a function of the preshock temperature for Abell~1142 as shown in Figure~\ref{fig:vt}.
The green solid line represents a postshock temperature ($T_2$) of 3.5 keV (Figure~\ref{fig:spec}) which is a projected temperature. The actually postshock temperature is more likely to be 4.0 keV (or even 4.5 keV), marked in magenta line. 
The actual scenario could be anywhere between a 1100\,km\,s$^{-1}$ merger in an originally 2.6 keV ICM and a 1550\,km\,s$^{-1}$ merger in an originally 1.4 keV ICM. Even these upper and lower limits are not hard limits since these values are just rough estimates. Considering that the ICM temperature of Abell~1142 drops to 2.5 keV out to 200 kpc (Figure~\ref{fig:spec}), we use 2.5 keV as the preshock temperature of the ICM. This corresponds to a merging velocity of 1200\,km\,s$^{-1}$ between the two subclusters.  

The cool core at the center of Abell~1142 could be a remnant of either G1 or G2 as the merger displaced this cool core from its BCG. Supposedly, the total velocity difference is 1200\,km\,s$^{-1}$ with a line of sight velocity difference of $1000$\,km\,s$^{-1}$. This corresponds to a velocity difference of $660$\,km\,s$^{-1}$ in the plane of the sky between G1 and G2 (with G1 moving eastward and G2 moving westward). G1 or G2 may move at a transverse velocity of $330$\,km\,s$^{-1}$ relative to the rest frame. 
It is unclear whether this ``cool core" used to belong to G1 or G2. In projection, it overlaps with member galaxies in the G1 subcluster and the radio tail of G1 pointing directly to this cool core. But it is more connected to G2 than G1 through X-ray emission and it is more likely for the less massive subcluster to have its cool core displaced in a merger. In either case, the current ``cool core" is 100 kpc away from its BCG in projection. The separation between the cool core and its BCG may have occurred 100 Myr ago.

\subsection{\bf Chemical enrichment processes}

The spectral analysis of Abell~1142 shows its ICM has a metal abundance peak of $\sim 0.7$\,Z$_{\odot}$ at the cluster center which is cooler than the ambient ICM.
The origin of the metals at the cluster center should be directly related to the formation of this cool region lacking a BCG. At least two pictures start to emerge: (1) this cool region was formed through the accumulation of cold gas stripped from infalling subgroups/galaxies during their passages of the cluster center, or (2) the major merger left a cool core (as in the cool core remnant scenario, Rossetti \& Molendi 2010) which was stripped off from its BCG.
In the first scenario, the source of the metal is supplied by the newly accreted/formed subgroups/galaxies; such metals should be rich in SNII production. In the second scenario, the source of the metal is the ejection from the BCG before the cool core was separated from its BCG; such metals should be rich in SNIa production. 
{Indeed, this metal rich cool region could be the ISM stripped from either G1 or G2 rather than being a cool core remnant. Typical examples include the ram pressure stripped galaxy M86 falling through the Virgo Cluster (Randall et al.\ 2008), the stripped subcluster as a result of the off-axis merger in RXJ0751+5012 (Russell et al.\ 2012), and the prominent stripped tail in the outskirt of Abell~2142 (Eckert et a.\ 2014). Then again, unlike these typical stripped tails, the cool region in Abell~1142 is not in a tail-like shape and it lacks uninterrupted bridge of emission from the main body.}
Deeper exposure of Abell~1142 would allow us to determine the metal content of elements other than Fe (e.g., $\alpha$-elements) with which we can unveil the enrichment process of Abell~1142 and cast light on the formation of this cool region.

The metal abundance profile of Abell~1142 gradually declines to $\sim$0.1\,Z$_{\odot}$ out to 200 kpc. 
This could be a result of ``Fe-bias" caused by multi-temperature gas (Buote 2000) and {its metallicity could still be as high as $\sim0.3$\,$Z_{\odot}$ due to the large uncertainty}. Nevertheless, a possible low abundance in the ICM outside the cluster core would suggest a lack of early enrichment in poor clusters. This would provide a sharp contrast to massive CC clusters like the Perseus Cluster with its ICM uniformly enriched ($\sim0.3$\,$Z_{\odot}$) out to the virial radius (Werner et al.\ 2013). 
Nevertheless, this may support that infalling galaxies and sub-groups in poor clusters may be able to retain their enriched gas until they reach the denser cluster core (Elkholy et al.\ 2015; Su et al.\ 2014).

\section {\bf Conclusions}
We present a {\sl Chandra} observation of Abell~1142, a poor cluster containing two BCGs (G1 and G2). Its central X-ray emission peak is offset from both BCGs by $\sim$100 kpc. 
This peak corresponds to a cool and metal enriched region compared with the surrounding cluster gas. 
The azimuthally-averaged surface brightness profile centered on this X-ray peak can be well described by a double $\beta$ model. 
These features are very suggestive of a cluster cool core.
The redshift distribution of its member galaxies indicates that Abell~1142 may consist of two subclusters that are merging at a relative velocity of $\approx$1000\,km\,s$^{-1}$ in the line-of-sight. G1 belongs to the more massive subcluster while G2 belongs to the less massive subcluster. 
Its ICM temperature may be elevated due to the shock heating. 

Perhaps, one of the subclusters used to be a typical relaxed CC cluster. This merger caused its metal-enriched cool core to be displaced from its BCG.
We may witness the onset of a CC cluster being transformed into a NCC cluster through major mergers. 
In an alternative scenario, this metal rich cool region is formed through the accumulation/stripping of cold gas supplied by infalling subgroups/galaxies.

We note that G1 is made up of three individual galaxies residing within a radius of 5 kpc in projection. Either the dynamical friction is smaller than expected from a cuspy dark matter halo or we happened to capture these galaxies as they are merging before they sink into cluster center.

\section{\bf Acknowledgments}
We would like to thank the helpful discussions with William Forman, Manoj Kaplinghat, Tiziana Venturi, and Ralph Kraft.
D.A.B. and Y.S. gratefully acknowledge 
partial support from the National
Aeronautics and Space Administration under Grant No.\ NNX13AF14G
issued through the Astrophysics Data Analysis Program. Partial support
for this work was also provided by NASA through Chandra Award Number
GO2-13159X issued by the Chandra X-ray Observatory Center, which is operated by the Smithsonian Astrophysical Observatory for and on behalf of NASA under contract NAS8-03060. 
This publication makes use of data products from the Two Micron All Sky Survey, which is a joint project of the University of Massachusetts and the Infrared Processing and Analysis Center/California Institute of Technology, funded by the National Aeronautics and Space Administration and the National Science Foundation. Funding for the SDSS and SDSS-II has been provided by the Alfred P. Sloan Foundation, the Participating Institutions, the National Science Foundation, the U.S. Department of Energy, the National Aeronautics and Space Administration, the Japanese Monbukagakusho, the Max Planck Society, and the Higher Education Funding Council for England. The SDSS Web Site is {\url http://www.sdss.org/}.
The SDSS is managed by the Astrophysical Research Consortium for the Participating Institutions.

\end{document}